\documentstyle[12pt]{article}

\advance\voffset by -2.0cm
\advance\hoffset by -3.0cm
\def\3{\ss}

\newcommand{\A}{{\cal A}}
\newcommand{\Ao}{{\cal A}_0}
\newcommand{\Au}{{\cal A}_0(\overline{\cal M})}
\newcommand{\Hi}{{\cal H}}
\newcommand{\Of}{{\cal O}}
\newcommand{\M}{{\cal M}}
\newcommand{\ro}{\varrho}
\newcommand{\Des}{\Delta(S_0)}

\newcommand{\li}{\begin{list}{(\roman{enumi})}{\usecounter{enumi}}}

\newcommand{\F}{\mbox{\boldmath ${\cal F}$}}

\newcommand{\npssvs}{scattering state vectors}
\newcommand{\tI}{\tilde{I}}
\input{epsf}

\def\sq{\hbox{\rlap{$\sqcap$}$\sqcup$}}
\def\qed{\ifmmode\sq\else{\unskip\nobreak\hfil
\penalty50\hskip1em\null\nobreak\hfil\sq
\parfillskip=0pt\finalhyphendemerits=0\endgraf}\fi}

\def\bbbr {{\rm I\!R}}
\def\bbbn {{\rm I\!N}}
\def\bbbone {{\mathchoice {\rm 1\mskip-4mu l} {\rm 1\mskip-4mu l}
{\rm 1\mskip-4.5mu l} {\rm 1\mskip-5mu l}}}
\def\bbbc{{\mathchoice {\setbox0=\hbox{$\displaystyle\rm C$}\hbox{\hbox
to0pt{\kern0.4\wd0\vrule height0.9\ht0\hss}\box0}}
{\setbox0=\hbox{$\textstyle\rm C$}\hbox{\hbox
to0pt{\kern0.4\wd0\vrule height0.9\ht0\hss}\box0}}
{\setbox0=\hbox{$\scriptstyle\rm C$}\hbox{\hbox
to0pt{\kern0.4\wd0\vrule height0.9\ht0\hss}\box0}}
{\setbox0=\hbox{$\scriptscriptstyle\rm C$}\hbox{\hbox
to0pt{\kern0.4\wd0\vrule height0.9\ht0\hss}\box0}}}}

\newtheorem{Theorem}{Theorem}[section]
\newtheorem{Lemma}[Theorem]{Lemma}
\newtheorem{Definition}[Theorem]{Definition}

\title{
Scattering States of Plektons (Particles with Braid Group
Statistics)  in $2+1$ Dimensional Quantum Field Theory}
\author{K. Fredenhagen
\thanks{email: i02fre@dsyibm.desy.de}\\
II. Institut f\"ur Theoretische Physik\\
Universit\"at Hamburg,
D 22761 Hamburg  \\
\\
M.R. Gaberdiel
\thanks{Partly supported by `Studienstiftung des deutschen Volkes'.}
\thanks{email: M.R.Gaberdiel@amtp.cam.ac.uk}\\
Department of Applied Mathematics
and Theoretical Physics,\\
University of Cambridge, Silver Street,
Cambridge, CB3 9EW, U.K. \\
\\
S.M. R\"uger\\
Informatik, Sekr. FR 5-9 \\
TU Berlin, D 10587 Berlin  }
\date{October 10, 1994}
\begin{document}
\maketitle
\vspace{-14.5cm}

\begin{flushright}
DAMTP-94-90 \\
hep-th/9410115
\end{flushright}
\vspace{14.5cm}

\begin{abstract}
A Haag-Ruelle scattering theory for particles with
braid group statistics is developed, and the arising structure
of the Hilbert space of multiparticle states is analyzed.
\end{abstract}

\section{Introduction}
Particles in $2+1$ dimensional spacetime are not necessarily
bosons or fermions; in general, their statistics may be described
by a unitary representation of Artin's braid group \cite{Artin}.
Such particles will be called plektons, in the following.
In a quantum mechanical framework the possible
existence of plektons in $2$ space dimensions was first
observed by Leinaas and Myrheim \cite{LM} in their analysis
of the principle of indistinguishability of identical
particles. In the framework of quantum field theory
the presupposed correspondence between particles and local
fields seemed to forbid exotic statistics in more than
$1+1$ dimensions. But adopting the point of view of algebraic
quantum field theory that locality has to be assumed only
for observables, Buchholz and one of us showed \cite{BuFre}
that even in purely massive theories particles might
correspond to non-local fields with the consequence
that ordinary statistics could be derived only in $3+1$
(or more) dimensions.

Models for particles with a one dimensional representation
of the braid group (``anyons") were first invented by
Wilczek \cite{W}. Non-abelian gauge theories with a Chern-Simons
term in the action are candidates for models with non-abelian
braid group statistics. Anyons are considered to be
the excitations which are responsible for the fractional
Quantum Hall Effect~\cite{Laughlin}.

In order to predict phenomena caused by plektons
a multiparticle formalism is desirable.
In the case of permutation group statistics the multiparticle
space (as a representation space of the Poincar\'e group)
is obtained by the choice of a Poincar\'e invariant metric
(determined by the statistics) on the tensor product
of Poincar\'e group representations on single particle spaces
(see \cite{DHR}). This is no longer true in the plektonic
case because the sum rules for spins involve the statistics
\cite{Fre1}. A multiplekton space which satisfies these
requirements was recently constructed by Mund and Schrader
\cite{MuSch}: it is determined by the Poincar\'e group representation
in the single particle spaces and a representation of the braid
group ${\rm B}_n$.

In this paper we show that in quantum field theory
with braid group statistics scattering states exist, and we
compute the structure of the ``plektonic Fock space" of
scattering states. This space turns out to be a direct
sum of the $n$-particle spaces of Mund and Schrader
where the braid group representations are induced by
a Markov trace on the braid group ${\rm B}_{\infty}$.
Some work in this direction
has already been done by Fr\"ohlich and Marchetti \cite{FM},
who concentrated on the abelian case,
and by Schroer \cite{Schroer}, who pointed out problems
and made some prospective remarks.

This work is based in parts on the diploma theses of two of us
\cite{Ruger,Gab}. In Section 2 we describe the general
framework as it was developed in the framework of
algebraic quantum field theory \cite{DHR}, \cite{Haag}. In
order to obtain a coherent description of the non-local
operators which one has to add to the algebra
of local observables we use the formalism of
\cite[II]{FRS} (see also \cite{Fre2}, \cite{Guido}) where an
extension $\Au$ of the algebra of local observables
was introduced which may be
considered as the algebra of local observables on the
union of Minkowski space with the hyperboloid at spacelike
infinity. We then define the associated field bundle \cite{DHR}
(an intrinsic structure equivalent to the exchange
algebra of vertex operators \cite{Rehren1} which is known
from conformal field theory) and compute the commutation
relations of generalized fields.

In Section 3 we construct Haag-Ruelle approximants to
scattering states. This can be done as in \cite{BuFre}
(see e.g. \cite{FM}). The computation of scalar
products is somewhat complicated and some formulae
guessed previously could not be confirmed. There is a
subtle point concerning the dependence of the scattering
states on the choice of the Lorentz frame. As pointed
out to us by Schrader some years ago, such a difficulty
had to be expected in view of the absence of a
Lorentz invariant ordering on a mass hyperboloid in 3 dimensional
Minkowski space,
and actually, the proof in \cite{BuFre} that the
construction is Lorentz
invariant, breaks down in $3$ dimensions.
We therefore reformulate the Haag-Ruelle theory in a
manifestly Lorentz invariant way. There remains a dependence
of the scattering vectors on the spacelike directions
which characterize the localization of the stringlike
localized fields by which the vectors were generated.

This set of directions might be considered as describing
a local
trivialization of a hermitian
vector bundle over the configuration space $C_n$ of
$n$ non-coincident
velocities in 3-dimensional Minkowski space with a
representation of the
pure braid group as structure group. This bundle
is associated to the principal bundle $p:\tilde{C}_n
\rightarrow C_n$ where $\tilde{C}_n$ is the universal
covering space of $C_n$ with projection $p$ and the pure
braid group as structure group \cite{St}.

This structure (for the case of single particle spaces
with an irreducible representation of the Poincar\'e group)
was anticipated by Schrader \cite{Schrader} (see also
\cite{Tscheuschner}) and further elaborated in \cite{MuSch}.

\section{Commutation relations of generalized local fields}
\renewcommand{\theequation}{2.\arabic{equation}}
Let us start with describing the algebraic framework of
quantum field theory
\footnote{For more details see \cite{DHR}, \cite{Haag} and references
therein.}:
We are given a family of von Neumann algebras $\A(\Of)$
(the algebras of observables measurable within $\Of$)
on some Hilbert space $\Hi_0$
indexed by the open double-cones $\Of$ in Minkowski space $\M$
which satisfy the following properties:
\begin{list}{(\roman{enumi})}{\usecounter{enumi}}
\item $\A(\Of_{1})\subset\A(\Of_{2})$ \hspace{0.5cm}if
 $\Of_1\subset\Of_2$ (isotony) \\
 We can then define the algebra of observables $\Ao(\M)$ to be
 the norm closure of the union $\bigcup\A(\Of)$ of these local algebras.
 Moreover, $\Ao({\cal R})$ for an arbitrary region ${\cal R} \subset \M$
 is defined as the C*-subalgebra of $\Ao(\M)$ generated
 by all algebras $\A(\Of)$ with double-cones $\Of \subset
 {\cal R}$, and $\A({\cal R})$ is its weak closure.
\item $\A(\Of_1)\subset\A(\Of_2)'$ \hspace{0.5cm} if
 $\Of_1\subset\Of_2'$ (locality)\\
 Here $\A(\Of_2)'$ is the commutant of $\A(\Of_2)$
 and $\Of_2'$ denotes the spacelike complement
 of $\Of_{2}$ in Minkowski space.
\item There is a representation $\alpha$ of the identity component
 ${\cal P}_{+}^{\uparrow}$
 of the Poincar\'{e} group by automorphisms of $\Ao(\M)$
 such that
 \begin{equation}
  \alpha_{(x,\Lambda)}(\A(\Of))=\A((x,\Lambda)(
  \Of)).
 \end{equation}
\item There is a
 strongly continuous unitary representation $U_0$ of
 ${\cal P}_{+}^{\uparrow}$ on $\Hi_0$ such that
 \begin{itemize}
 \item $U_0(L)AU_0(L)^*=\alpha_L(A)$ for all $L\in {\cal P}_+^\uparrow$,
      $A\in \A_0(\M)$.
 \item The generators of the translations $P_{\nu}$ satisfy the
  spectrum condition
  \begin{equation}
   {\rm sp} P \subset \{0\}\cup\{p\in \M\;|\;
   p^{2}>\mu^{2},p_0>0\}
  \end{equation}
  for some $\mu\geq 0$.
 \item There is a cyclic unit vector $\Omega\in\Hi_0$, unique up to a
 phase, such that $U_0(x,\Lambda)\Omega=\Omega$.\\ $\Omega$ represents
 the vacuum.
 \end{itemize}
\item Haag duality for spacelike cones (see below).
\end{list}
The embedding of $\A_0(\M)$ into the algebra
${\cal B}(\Hi_0)$ of all bounded operators on $\Hi_0$
is called the vacuum representation $\pi_0$.

In the present paper we want to analyze multiparticle scattering
states in a purely massive theory.
We are therefore interested in  representations of the
algebra of observables which describe massive particles, i.e.\
representations
$\pi$ of $\Ao(\M)$ by bounded operators on a Hilbert space $\Hi_{\pi}$
together with a strongly continuous representation $U_{\pi}$
of the covering group $\widetilde{P_{+}^{\uparrow}}$ of
$P_{+}^{\uparrow}$ satisfying
\begin{equation}
\label{Poincare}
 {\rm Ad} U_{\pi}(g)\circ\pi=\pi\circ\alpha_{L(g)}
\end{equation}
for any $g\in \widetilde{P_{+}^{\uparrow}}$, where
$g\mapsto L(g)$ is the covering homomorphism and
the generators of the translations are required to fulfill the
spectrum condition
\begin{equation}
\label{spectrum}
 H_m\subset {\rm sp} P \subset
 H_m\cup\{p\in {\cal M}\;|\;p^2>M^2, p_0>0\}
\end{equation}
with $0<m<M$. Here $H_m$ is the mass shell
$H_m=\{p\in {\cal M}\;|\;p^2=m^2, p_0>0\}$, $m$
is interpreted as the mass of the particle described by $\pi$
and $\pi$ is called ``massive single particle representation".

It was shown in \cite{BuFre} that for irreducible massive
single particle representations
$\pi$ there is a unique vacuum representation $\pi_0$,
i.e.\ a representation satisfying (iv) (with $\mu\geq M-m$)
such that
$\pi$ and $\pi_0$ are unitarily equivalent when restricted to the
algebra of the causal complement of any spacelike cone $S$
\begin{equation}
\label{equiv}
 \pi\left|_{\Ao(S')}\right.\cong\pi_0\left|_{\Ao(S')}\right. .
\end{equation}
A  spacelike cone $S$ is the convex set
\begin{equation}
 S:=a+\bigcup_{\lambda>0} \lambda\Of ,
\end{equation}
where $a\in \M$ is the apex
and $\Of$ is a double-cone of
spacelike directions
\begin{equation}
 \Of=\{r\in \M|r^{2}=-1\; \mbox{and}\; r_{+}-r, r-r_{-}\in V_{+}\}
\end{equation}
with $r_{+}^{2}=r_{-}^{2}=-1$,  $r_{+}-r_{-}\in V_{+}$, $V_+$
denoting the interior of the forward light cone.
We denote the set of spacelike cones by ${\cal S}$.

In view of this result we shall from now on fix the
vacuum representation $\pi_0$ and identify it with the
defining (identical) representation of $\A_0(\M)$ on
$\Hi_0$. We consider only those
massive single particle representations $\pi$ which satisfy
the ``selection criterion" (\ref{equiv}) with respect to $\pi_0$.
Furthermore, we shall assume that the fixed vacuum representation
fulfills Haag duality for spacelike cones, i.e.\
\begin{equation}
 \A(S')=\A(S)'
 \hspace{1.0cm} \mbox{for all $S\in{\cal S}$.}
\end{equation}

One now proceeds as in the DHR-analysis \cite{DHR} and identifies
the representation spaces $\Hi_0$ and $\Hi_{\pi}$. To this end
one exploits (\ref{equiv}) to define
for any representation $\pi$, satisfying (\ref{equiv}),
corresponding homomorphisms
$\ro:\Ao(\M)\rightarrow B(\Hi_{0})$
which are unitarily equivalent to $\pi$. In the present
context these $\ro$ are in general not endomorphisms,
i.e.\ $\ro(\Ao(\M))\not\subset\Ao(\M)$, and thus the
usual definition for the composition of sectors is ill-defined
\cite{BuFre}.

To overcome this difficulty one can pass from the algebra
of quasi-local observables $\Ao(\M)$ to a larger $C^{*}$-algebra $B^{r}$
depending on a forbidden spacelike direction $r$
\begin{equation}
 B^{r}:=\overline{\bigcup_{S\in{\cal S}(r)} \A(S')} .
\end{equation}
Here $r$ is a spacelike unit vector (i.e.\ $r^2=-1$),
                          and ${\cal S}(r)$ is the set
of spacelike cones, which ``contain the direction $r$"
\begin{equation}
 {\cal S}(r):=\{S\in{\cal S}|\bar{S}+r\subset S\}.
\end{equation}
One can subsequently show (see \cite{BuFre} resp. \cite{Ruger}
for details) that if $\ro$ is a morphism localized
in a cone $S$ spacelike to $r$ (i.e.\ there exists a
$\hat{S}\in{\cal S}(r)$ such that $\hat{S}\subset S'$) then
$\ro$ extends uniquely to an endomorphism $\ro^{r}$ of
$B^{r}$ which is weakly continuous on all $\A(S')$ for
$S\in{\cal S}(r)$.
Thus one can proceed in very much the same
way as in the DHR-analysis; in particular, one defines the
composition of sectors via the composition of the
corresponding morphisms on $B^{r}$. However, one has to check
that all structural properties are independent of the choice of
the direction~$r$.\\
To avoid singling out this 'auxiliary direction' one can embed
all $B^{r}$ into an even larger $C^{*}$-algebra, the {\em universal
algebra} $\Au$, which can be uniquely characterized by the
following universality conditions (this construction
was proposed in \cite{Fre2} and further developed
in \cite{Guido} and \cite[II]{FRS}):
\begin{itemize}
\item there are unital embeddings $i^{I}:\A(I)\rightarrow
 \Au$ such that for all $I,J\in {\cal K}:=
 \{S,S'|S\in{\cal S}\}$
 \begin{equation}
  i^{J}\left|_{\A(I)}\right.=i^{I}\hspace{1.0cm}
  \mbox{if $I\subset J$}
 \end{equation}
 and $\Au$ is generated by the algebras $i^{I}(\A(I))$.
\item for every family of normal representations $(\pi^{I})_{I\in
 {\cal K}}$,
 $\pi^{I}:\A(I)\rightarrow B(\Hi_{\pi})$ which satisfies
 the compatibility condition
 \begin{equation}
  \pi^{J}\left|_{\A(I)}\right.=\pi^{I} \hspace{1.0cm}
  \mbox{if $I\subset J$},
 \end{equation}
 there is a unique representation $\pi$ of $\Au$ in $\Hi_{\pi}$
 such that
 \begin{equation}
  \pi\circ i^{I}=\pi^{I}.
 \end{equation}
\end{itemize}
The usefulness of this definition
is due to
the fact that the
endomorphisms $\ro^{r}$ have a common extension to
an endomorphism $\ro$ of $\Au$
such that the unique extensions of $\pi$ and $\pi_{0}$ to $\Au$
(which shall be denoted by the same symbols) satisfy
\footnote{From now on we shall consider $\A(I), I\in
{\cal K}$ as abstract subalgebras of $\Au$ and only
$\pi(A)$ (resp. $\pi_{0}(A)$) as operators on the vacuum
Hilbert space $\Hi_0$.}
\begin{equation}
 \pi=\pi_0\circ\ro .
\end{equation}
However, in general the vacuum representation $\pi_0$
is no longer faithful
on $\Au$ (see \cite[II]{FRS} for details).\\
\medskip

The endomorphisms $\ro$ one obtains are localized in some
$I \in {\cal K}$ in the following sense \cite{FKy}.
\begin{Definition}
 An endomorphism $\ro$ of $\Au$ is called localizable
 within $I\in {\cal K}$ if for all $I_0 \subset I$, $I_0 \in {\cal K}$
 there exists a unitary $U \in \A(I)$ such that
 $$
  \ro(A)={\rm Ad}U (A) ,\quad A\in \A(I_0'),
 $$
 $$
  {\rm Ad}U^*\circ\ro(\A(I_1))\subset \A(I_1) ,\quad
  I_1 \supset I_0, I_1 \in {\cal K}.
 $$
 An endomorphism $\ro$ is called transportable if for every $I\in
 {\cal K}$ there exists on endomorphism $\ro'$ of $\Au$ which is
 localizable within $I$ and is inner equivalent to $\ro$, i.e.\
 there exists a unitary $U\in\Au$ such that $\ro'={\rm
 Ad}U\circ\ro$.
\end{Definition}
Note that endomorphisms which are localizable within $I$ are not
necessarily localizable within $J\supset I$. However transportable
endomorphisms which are localizable within some region are
automatically localized in every larger region.

Let $\Delta$ denote the set of transportable endomorphisms
and $\Delta(I)$ the subset of transportable endomorphisms
which are localizable within $I$.

In the $s+1$-dimensional situation, $s\geq 3$, it is possible to embed
$\Ao(\M)$ into a net of field algebras ${\cal F}$ which transform
covariantly under some compact group of internal symmetries and
satisfy graded locality \cite{DR}. These fields may generate
single particle states from the vacuum, and one can use them
for the construction of multiparticle scattering states by the
standard recipe of the Haag-Ruelle theory \cite{HR,Haag}.
At the time when \cite{DHR} and \cite{BuFre} were written the existence
of field algebras was not established, so a different method for
the construction of scattering states had to be used. This method
is based on the fact that the partial intertwiners which exist
between representations satisfying a localizability condition
of the type (\ref{equiv}) behave in many respects in the same way as
field operators. A convenient formalism for the description
of these partial intertwiners is the so-called field bundle
introduced in \cite[II]{DHR}. In cases where nontrivial
braid group statistics occurs a general construction of field
algebras is difficult (see, however, \cite{MS,Rehren2})
Interestingly
enough, the exchange algebra formalism of chiral conformal
field theory \cite{Rehren1} is equivalent to the field bundle
formalism \cite[I]{FRS}.

In our case, the field bundle is defined as follows. We fix
a spacelike cone $S_0$.

\begin{list}{(\roman{enumi})}{\usecounter{enumi}}
\item  We describe vectors $\Psi$ in some representation $\pi_0
 \circ\ro$ by a pair $ {\bf \Psi}=\{\ro;\Psi\}$ and consider
 $\Delta(S_0)\times\Hi_0 =
 {\bf \Hi} $ as a hermitian vector bundle over $\Delta(S_0)$,
 where on every fiber $\Hi_{\ro} =\{\ro\}\times\Hi_0$
 the scalar product is that of $\Hi_0$.
\item {\em Generalized field operators} are
 pairs $ {\bf B}=\{\ro;B\}\in\Des\times\Au $.
 They act on ${\bf \Hi}$ by
 \begin{equation}
  \{\hat{\ro};B\}\{\ro;\Psi\}=\{\ro\hat{\ro};
  \pi_0\circ\ro(B)\Psi\}
 \end{equation}
 and possess the norm
 $\|\{\hat{\ro};B\}\|:=\|B\|. $
 The fibers $\{\ro\}\times\Au$ are linear spaces isomorphic
 to $\Au$.
 In addition, there is an
 associative multiplication law of field operators
 (consistent with the above action on ${\bf \Hi}$)
 \begin{equation}
  \{\ro_1;B_1\}\{\ro_2;B_2\}:=\{\ro_2\ro_1;
  \ro_2(B_1)B_2\}.
 \end{equation}
 Observables correspond to field operators of the form
 $\{id;A\}, A\in\Au$.
\item An element $T$ of the global algebra $\Au$ with the
 property
 $ \ro_1(A)T=T\ro(A),\; A\in\Au $
 for two morphisms $\ro_{1}$ and $\ro$ is called an
 {\em intertwiner} from $\ro$ to $\ro_1$ . It induces
 the actions
 \begin{equation}
\label{intert}
  (\ro_1|T|\ro)\{\ro;\Psi\}=\{\ro_1;\pi_0(T)\Psi\},
  \hspace{1.5cm}
  (\ro_1|T|\ro)\circ\{\ro;B\}=\{\ro_1;TB\}.
 \end{equation}
\item Poincar\'{e} transformations are implemented in the field
 bundle in the following way. Let
$U_{\ro}(g)$ be the representation of $\widetilde{
P_{+}^{\uparrow}}$ corresponding to $\pi=\pi_0\circ\ro$
(see (\ref{Poincare})) then
\begin{eqnarray}
{\displaystyle  U(g)\{\ro;\Psi\}} =
{\displaystyle \{\ro;U_{\ro}(g)\Psi\}} \\
{\displaystyle  \alpha(g)\{\ro;B\}} =
{\displaystyle \{\ro;Y_{\ro}(g)  \alpha_{L(g)}B\},}
\label{transform}
\end{eqnarray}
where $\alpha_{L}$ denotes the extension of $\alpha_{L}$ from
$\Ao(\M)$ to $\Au$ and
\begin{equation}
 Y_{\ro}(g)=\pi_0^{-1}(U_{\ro}( g)\;U_0(L(g))^{-1}).
\end{equation}
(\ref{transform}) is to be understood only for $g$ sufficiently close to
the identity in order to make sure that there is a path $L(t)$ in the
homotopy class $g$ such that
$\bigcup_t L(t)S_0$ is spacelike to some spacelike direction $r$.
Then $U_{\ro}(g)\;U_0(L(g))^{-1}\in\pi_0(B^{r})$.
This guarantees that the preimage under $\pi_0$ is well-defined
since $\pi_0$ is faithful on $B^{r}$. The general
case is obtained by successive use of (\ref{transform}).\\
Poincar\'{e} transformations commute with intertwiners, i.e.\
if $T$ is an intertwiner from $\ro$ to $\ro_{1}$, then
\begin{equation}
\pi_{0}(T)\;U_{\ro}(g)=
U_{\ro_{1}}(g)\;\pi_{0}(T).
\end{equation}
Finally, the representation in the fiber $\ro_{1}\ro_{2}$
is related to the ones in the fibers $\ro_{1}$ and $\ro_{2}$
by
\begin{equation}
U_{\ro_{1}\ro_{2}}(g)=\pi_{0}\circ\ro_{1}
\left(Y_{\ro_{2}}(g)\right)\;U_{\ro_{2}}(g)
\end{equation}
(see \cite[II]{FRS} for details).
\item  A necessary condition for a generalized field operator
${\bf B}=\{\ro;B\}$ to be localized in $I\in {\cal K}$
is that $B$ intertwines the identity with $\ro$ on
$\A(I')$. But due to the existence of global self-intertwiners,
the intertwining property of $B$ is too weak for a derivation
of commutation relations between spacelike separated generalized
fields. Instead one characterizes the localization by a path
in ${\cal K}$, i.e.\ a finite sequence $I_i\in {\cal K}$,
$i=0,\dots,n$ with $I_0=S_0$ and such that either $I_i\subset I_{i-1}$
or $I_i\supset I_{i-1}$, $i=1,\dots,n$. For each $i$ there is
some unitary $U_i\in \A(I_i\cup I_{i-1})$ such that
${\rm Ad}U_i\cdots U_1\circ\ro\in\Delta(I_i)$. Then $\{\ro,B\}$
is called localized in $(I_0,\dots,I_n)$ if
\begin{equation}
 U_n\cdots U_1 B \in \A(I_n).
\end{equation}
\end{list}
The concept of localization described above is an extension of the
corresponding notion in \cite{DHR} following ideas of \cite[II]{FRS}.
Clearly, the localization depends only on the homotopy class
$\tI$ of a path $(I_0,\dots,I_n)$ where homotopy is
defined in the obvious way. The set of these classes shall be
denoted by $\tilde{\cal K}$ and the set of field operators
localized in $\tI$ by $\F(\tI)$.

Let us now consider paths with the same endpoint. They differ
(up to homotopy) by a closed path $\gamma = (I_0,\dots,I_k)$
with $I_k = I_0$.
We choose associated intertwiners $U_1,\dots,U_k$
with $\pi_0(U_k\cdots U_1) = 1$. Then $\gamma \mapsto
U(\gamma)=U_k\cdots U_1$ is a representation of the homotopy
group by unitary elements of $\Au$.

In a next step we look at products of field operators with
mutually spacelike localization. Here $(I_0,\dots,I_n)=\tilde{I}$
and $(J_0,\dots,J_k)=\tilde{J}$, $I_0=J_0=S_0$,
are called mutually spacelike if the endpoint of $\tilde{I}$,
$e(\tilde{I})=I_{n}$, is spacelike separated from $e(\tilde{J}) = J_{k}$.
Let ${\bf B}_i =\{\ro_i,B_i\}$ be localized in $\tilde{I_i}$,
$i=1,\dots,n$. Then
\begin{equation}
{\bf B}_{\sigma(n)} \cdots {\bf B}_{\sigma(1)} =
\varepsilon \circ {\bf B}_n \cdots {\bf B}_1 ,
\end{equation}
where $\varepsilon$ is an intertwiner from $\ro_1 \cdots
\ro_n$
to $\ro_{\sigma(1)} \cdots \ro_{\sigma(n)}$. $\varepsilon$
depends on the endomorphisms $\ro_i \in \Des$, on the
localizations $\tI_i$ and on the permutation $\sigma$.
It is described in terms of a unitary representation of the
groupoid of colored braids on the cylinder \cite[II]{FRS}.
The associated braid can be obtained by the following
geometrical construction.

Mutually spacelike paths
$\tI_i$ are continuously deformed to
paths $\gamma_i$ on the set of spatial directions in some
Lorentz frame, i.e.\ to paths on the circle
$\rm{S}^1$ with a fixed initial point $z_0$ corresponding to
$I_0$ and disjoint
endpoints $z_i$ corresponding to the endpoints $e(\tI_i)$
of $\tI_i$.
On the cylinder ${\rm S}^1 \times \bbbr$ we
choose points $(z_0,i)$, $i=1,\dots,n$
and paths $\Gamma_i$ from $(z_0,i)$ to $(z_0,\sigma(i))$,
\begin{equation}
 \Gamma_i =(\gamma_i^{-1},\sigma(i)) \circ (z_i,i\rightarrow
 \sigma(i)) \circ (\gamma_i,i).
\end{equation}
The braid is now the usual equivalence class of the family of
strands $\Gamma_i, i=1,\dots,n$ (see for example figure~1,
where the 3rd dimension is introduced for visualizing
the parameter of the paths $(z_i,i\rightarrow\sigma(i))$).

\begin{figure}
\hspace*{5.25cm}
\epsfxsize=5.5cm
\epsfbox{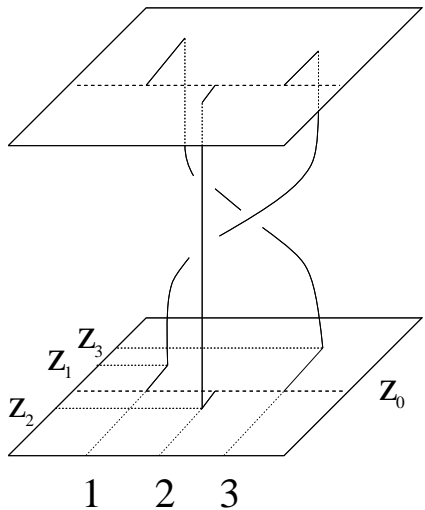}
\caption{The braid corresponding  to the
permutation $\sigma = \tau_{1} \tau_{2} \tau_{1}$ in the special
case where all three paths $\gamma_{i}$ have trivial winding number
and thus can be represented as paths in the plane.}
\end{figure}

By the standard techniques of algebraic field theory
(see \cite{DHR,FRS} for more details) it follows that $\varepsilon$
is invariant under small deformations of
$\tI_1,\dots, \tI_n$ ---
so equivalent families $\Gamma_i, i=1,\dots,n$
give the same intertwiner $\varepsilon$ ---
and that the braid relations
are respected.

For the calculation of scalar products of scattering state vectors
in the next section we need the notion of a {\it left inverse}
of an endomorphism $\ro \in \Des$
(see \cite{FRS}, \cite{Fre4},
\cite{Longo} for more details):
A left inverse $\phi$ of a $\ro$ is
a positive mapping of $\Au$ mapping $\A(S_0)$ into itself
such that $\phi\circ\ro=id$ and such that
$\ro\circ\phi$ is a conditional expectation from
$\Au$ to $\ro(\Au)$. If $\ro$ is irreducible, i.e.
$\pi\circ\ro$ is irreducible, and has finite statistics,
a property which is automatically satisfied for irreducible
single particle representations \cite{Fre4}, the left inverse
is unique. If $\ro_1,\dots,\ro_n\in\Des$ are irreducible with
finite statistics, the product $\ro=\ro_1\cdots\ro_n$
is not irreducible, in general, and there is no unique
left inverse. But there is a so called standard left inverse
which is given by
\begin{equation}
 \phi=\phi_n\cdots\phi_1
\end{equation}
with $\phi_i$ the unique left inverse of $\ro_i,i=1,\dots,n$.
The standard left inverse is a trace on the algebra of local
self-intertwiners of $\ro$, i.e. if $S,T\in \A(S_0)$
commute with $\ro(\Au)$ then
\begin{equation}
 \phi(ST)=\phi(TS).
\end{equation}

Some of the following formulae are more easily expressed in terms
of {\it right inverses} of endomorphisms which have been recently
introduced by Roberts \cite{Roberts}. The right inverse of $\ro$
is only defined on the class of intertwiners of the form
$(\ro''\ro | T | \ro'\ro)$. For such an intertwiner,
the right inverse, $\chi_{\ro}(T)$, is an intertwiner
from $\ro'$ to $\ro''$. If $\ro$ has a conjugate representation,
a right inverse of $\ro$ can be defined as
\begin{equation}
\label{rightinv}
\chi_{\ro}(T) = \ro ''(\bar{R})^{*} \; T \; \ro'(\bar{R}),
\end{equation}
where $\bar{R}$ is an isometric intertwiner from the vacuum
representation to $\ro \bar{\ro}$.
Roberts has shown that there is a unique right inverse,
the {\it standard right inverse}, which agrees with
the standard left inverse on local self-intertwiners. The
standard right inverse is unique for  irreducible $\ro$ with
finite statistics. Furthermore, the product of standard right
inverses is the standard right inverse of the composite endomorphism.

We also need a version of the cluster theorem \cite{DHR}
which is adapted to the present situation.
Let ${\bf B}_i =\{\ro_i,B_i\}\in \F(\tI_i), i=2,4$
with $\tI_2=\tI_4$ and let ${\bf B}_j=\{\ro_j,B_j\}, j=1,3$ be products
of field operators. For fixed $e$ , $e^{2}=1$ let $\tau$ be the
supremum of $|t|$ for all $t$ for which all the field operators in
${\bf B}_1$ and ${\bf B}_{3}$ are spacelike localized with respect to
$I_{2} + t e $. Furthermore, let $T$
be an intertwiner from $\ro_1\ro_2$
to $\ro_3\ro_4$. We are interested in the leading behavior of
$({\bf B}_4{\bf B}_3{\bf \Omega},T {\bf B}_2{\bf B}_1{\bf \Omega})$ for
large $\tau$.
Let us assume that $\ro_4$ is irreducible
with finite statistics with right inverse $\chi_4$ and denote by $\{W_j\}$
a (possibly empty) orthonormal basis of the Hilbert space
of local intertwiners from $\ro_4$ to $\ro_2$.
Then we have the following
\begin{Lemma}
\begin{equation}
\label{Lemma2}
\left| ({\bf B}_4{\bf B}_3{\bf \Omega}, T {\bf B}_2{\bf B}_1
{\bf \Omega})- \sum_j
({\bf B}_3{\bf \Omega},\chi_4(T \ro_{1}(W_j)) {\bf B}_1{\bf \Omega})
({\bf B}_4{\bf \Omega},W_j^*{\bf B}_2{\bf \Omega}) \right|
\leq e^{-\mu\tau} \prod_i| |{\bf B}_i|| .
\end{equation}
\end{Lemma}

\noindent {\it Proof.} The proof of this lemma is similar to the proof
of Lemma~7.3 in \cite[II]{DHR}. However, introducing the concept of
a right inverse makes the present proof in our opinion conceptually
much clearer. In particular, it demonstrates that the proof
extends directly to the case of non-trivial statistics.

The field bundle is invariant (up to isomorphism)
under changing the location of $S_{0}$ locally. Therefore the same is true
globally, where, however, the isomorphism depends on the homotopy class
of the path connecting the two different localizations of $S_{0}$.
Using this property, we can assume without loss of generality that
the localization of $\tI_2=\tI_4$ coincides with $S_{0}$.
We calculate
\begin{equation}
\label{Rech1}
\begin{array}{rcl}
{\displaystyle
\left({\bf B}_4{\bf B}_3{\bf \Omega}, T {\bf B}_2{\bf B}_1
{\bf \Omega}\right)} & = &
{\displaystyle \left( \left(\bbbone_{\ro_{3}} \times \bar{R}_{4} \right)
{\bf B}_3{\bf \Omega}, {\bf B}_{4}^{\dagger} T {\bf B}_2{\bf B}_1
{\bf \Omega}\right)} \\
& = & {\displaystyle \left({\bf B}_3{\bf \Omega},
\left(\bbbone_{\ro_{3}} \times \bar{R}_{4}^{*} \right)
\left( T \times \bbbone_{\bar{\ro}_{4}} \right) {\bf B}_{4}^{\dagger}
{\bf B}_2{\bf B}_1 {\bf \Omega}\right)} \\
& = & {\displaystyle \left({\bf B}_3{\bf \Omega},
\left(\bbbone_{\ro_{3}} \times \bar{R}_{4}^{*} \right)
\left( T \times \bbbone_{\bar{\ro}_{4}} \right) \varepsilon (\ro_{2}
\bar{\ro}_{4}, \ro_{1} ) {\bf B}_1 {\bf B}_{4}^{\dagger}
{\bf B}_2 {\bf \Omega}\right)},
\end{array}
\end{equation}
where $\bar{R}_{4}$ is the local isometric intertwiner
from the vacuum to $\ro_{4} \bar{\ro}_{4}$ which appears in
the definition of the adjoint
operator in the field bundle formalism and $\varepsilon (\ro_{2}
\bar{\ro}_{4}, \ro_{1} )$ depends on the localization of ${\bf B}_{1}$
relative to $S_{0}$. (Here we use the usual notation for the intertwiner
calculus in the field bundle, see e.g.\ \cite{DHR} for definitions.)

\noindent The usual cluster theorem (see for example
\cite{Fre5} for a proof) implies that the scalar product is dominated
by the contribution of the vacuum sector in
${\bf B}_{4}^{\dagger} {\bf B}_2 {\bf \Omega}$. Inserting the
corresponding projector onto the vacuum
\begin{equation}
\sum_{j} \left|
(W_j\times\bbbone_{\bar{\ro}_4})
\bar{R}_{4} {\bf \Omega} \right\rangle \;
\left\langle
(W_j\times\bbbone_{\bar{\ro}_4})
\bar{R}_{4} {\bf \Omega} \right|
\end{equation}
into the scalar product,
we can conclude that the right hand side of (\ref{Rech1}) is
\begin{equation}
\sum_{j}
\left({\bf B}_3{\bf \Omega},
\left(\bbbone_{\ro_{3}} \times \bar{R}_{4}^{*} \right)
\left( T \ro_{1}(W_{j}) \times \bbbone_{\bar{\ro}_{4}} \right)
\varepsilon (\ro_{4} \bar{\ro}_{4}, \ro_{1} )
{\bf B}_1 \bar{R}_{4} {\bf \Omega} \right) \;
\left( \bar{R}_{4} {\bf \Omega}, {\bf B}_{4}^{\dagger} W_{j}^{*}
{\bf B}_2 {\bf \Omega}\right)
\end{equation}
up to a term which is smaller than the right hand side of (\ref{Lemma2}).
We observe that the second term in the product (for each $j$) is just
\begin{equation}
\left( \bar{R}_{4} {\bf \Omega}, {\bf B}_{4}^{\dagger} W_{j}^{*}
{\bf B}_2 {\bf \Omega}\right) =
\left({\bf B}_{4} {\bf \Omega}, W_{j}^{*} {\bf B}_2 {\bf \Omega}\right).
\end{equation}
We can commute the intertwiner $\bar{R}_{4}$ past ${\bf B}_1$ in the
first term in the product, and thus obtain (for each $j$)
\begin{equation}
\left({\bf B}_3{\bf \Omega},
\left(\bbbone_{\ro_{3}} \times \bar{R}_{4}^{*} \right)
\left( T \ro_{1}(W_{j}) \times \bbbone_{\bar{\ro}_{4}} \right)
\varepsilon (\ro_{4} \bar{\ro}_{4}, \ro_{1} )
\left( \bbbone_{\ro_{1}} \times \bar{R}_{4} \right)
{\bf B}_1  {\bf \Omega} \right).
\end{equation}
It remains to show that the product of intertwiners in this expression is
$\chi_{4}( T \ro_{1}(W_{j}))$. As the localization of
$\tI_2=\tI_4$ is $S_{0}$,
we can write the $\varepsilon$ intertwiner as
\begin{equation}
\varepsilon (\ro_{4} \bar{\ro}_{4}, \ro_{1} ) = V_{1}^{-1} \;
\ro_{4} \bar{\ro}_{4} (V_{1}),
\end{equation}
where $V_{1}$ is an intertwiner from $\ro_{1}$, localized in $S_{0}$, to
$\hat{\ro}_{1}$ whose localization corresponds to the localization
of the different operators in the operator product ${\bf B}_{1}$.
Thus (cf.\ \cite[II (2.21)]{FRS})
\begin{equation}
\begin{array}{rcl}
{\displaystyle
\varepsilon (\ro_{4} \bar{\ro}_{4}, \ro_{1} ) \bar{R}_{4}} & = &
{\displaystyle V_{1}^{-1} \ro_{4} \bar{\ro}_{4} (V_{1}) \bar{R}_{4}} \\
& = & {\displaystyle V_{1}^{-1} \bar{R}_{4} V_{1}} \\
& = & {\displaystyle V_{1}^{-1} \hat{\ro}_{1}(\bar{R}_{4}) V_{1}} \\
& = & {\displaystyle \ro_{1} (\bar{R}_{4}) ,}
\end{array}
\end{equation}
where, in the third line, we have used that $\bar{R}_{4}$ is local and
thus that $\hat{\ro}_{1}(\bar{R}_{4}) = \bar{R}_{4}$. By
(\ref{rightinv}) the product of intertwiners is indeed just
$\chi_{4}( T \ro_{1}(W_{j}))$.
\qed

\section{The structure of scattering states}

\renewcommand{\theequation}{3.\arabic{equation}}
\setcounter{equation}{0}
To construct multiparticle scattering states one wants to
follow the general recipe of the
Haag-Ruelle theory (for an introduction
see \cite{Haag}): one first constructs almost local
one particle creation operators ${\bf B}_i$ (here almost
localized in spacelike cones) and propagates them to other
times by using the Klein Gordon equation. In this way
one obtains operators ${\bf B}_i(t)$ which are essentially
localized at time $t$ and create one particle state vectors
${\bf \Psi}_i = {\bf B}_i(t) {\bf \Omega}$
independent of t. Then one proves convergence of
\begin{equation}
 {\bf B}_n(t)\cdots {\bf B}_1(t) {\bf \Omega}
\end{equation}
for $t \rightarrow \pm \infty$ and interprets the limits
as the outgoing or incoming, respectively, \npssvs\  corresponding
to single particle vectors ${\bf \Psi}_i,i=1,\dots n$.

The scalar products of \npssvs\ can be computed by using the
cluster theorem. In the case of finitely localized sectors
one can finally show that the scattering vectors do not
depend on the choice of the operators ${\bf B}_i(t)$ nor on
the choice of the time direction but only on the
single particle vectors ${\bf \Psi}_i,i=1,\dots,n$.
In the case of sectors localized in spacelike cones this
remains true in $s+1$ dimensional spacetime with $s>2$
\cite{BuFre}, but it definitely fails for $s=2$ if the sectors have
nontrivial braid group statistics.

This is the reason why the multiparticle space, considered as
a representation space of the Poincar\'e group, is in general
not isomorphic to a tensor product of one particle spaces
\cite{Fre1}. A satisfactory description of the scattering
space in terms of asymptotic fields does not yet exist.
However, the structure of this space can be completely computed and
turns out to be identical to the structure proposed by
Mund and Schrader \cite{MuSch}.

In a first step we show that a manifestly Lorentz invariant
formulation of the Haag Ruelle theory is possible. In this
formulation each particle propagates in its own rest frame.

Let ${\bf B}\in \F(\tI)$ for some localization
$\tI\in\tilde{\cal K}$
be such that the energy momentum spectrum
${\rm sp}_{\bf U}{\bf B}{\bf \Omega}$
contains an isolated mass shell $H_m$. Let
$f\in {\cal S}({\cal M})$ have a Fourier transform $\tilde{f}$ with
compact support in $V_+$ such that
${\rm supp}\tilde{f}\cap {\rm sp}_{\bf U}{\bf B}{\bf \Omega}
\subset H_m$.
Then we define
\begin{equation}
\label{creation}
 {\bf B}(t):=\int d^3x \;f_t(x)\;\alpha_x({\bf B}),
\end{equation}
where
\begin{equation}
 f_t(x)=(2\pi)^{-\frac{3}{2}}\;\int d^3p\; e^{-ipx+i(\frac{p^2-m^2}
{2m})t} \;\tilde{f}(p).
\end{equation}
${\bf B}(t){\bf \Omega}={\bf \Psi}$ is an eigenvector of the mass
operator $M^2=P^2$ with eigenvalue $m^2$ and it does not
depend on $t$.

\noindent The functions $f_t$ have the following properties:
\begin{Lemma}
Let $f\in {\cal S}({\cal M})$ have a Fourier transform $\tilde{f}$ with
compact support. Then the following two statements hold:
\begin{list}{(\roman{enumi})}{\usecounter{enumi}}
\item There exists a constant $c>0$ such that
$$\int d^3x\;|f_t(x)|<c\left( 1 + |t|^3  \right). $$
\item For each $\epsilon >0$, $N\in \bbbn$ there exists a constant
$C>0$ such that
$$|f_t(x)|<C\left|{\rm dist}(x,\frac{p}{m}t)+1\right|^{-N}$$
for all $p\in {\rm supp}\tilde{f}$ and all $x\in {\cal M}$, $t\in\bbbr$
such that ${\rm dist}(\frac{x}{t},\frac{q}{m})>\epsilon \;\;\forall q\in
{\rm supp}\tilde{f}$.
Here {\rm dist} means an Euclidean metric on $\M$.
\end{list}
\end{Lemma}
The proof follows by standard techniques of the stationary
phase approximation (see e.g. \cite{ReedSimon}).
\medskip

Let $V_\epsilon(f)=\{v\in \M, {\rm dist}(v,\frac{p}{m})<
\epsilon$ for some $p\in {\rm supp}\tilde{f}\}$.
It follows from the lemma that the operators ${\bf B}(t)$ can
be approximated by operators
${\bf B}_\epsilon(t)\in \F(\tI+tV_\epsilon(f))$,
\begin{equation}
 {\bf B}_\epsilon(t)=\int_{tV_\epsilon(f)}d^3x\;f_t(x)\;
 \alpha_x({\bf B}),
\end{equation}
such that $||{\bf B}(t)-{\bf B}_\epsilon(t)||<c_N|t|^{-N}$
for suitable constants $c_N$. Moreover, the norms of these operators
are bounded by $||{\bf B}(t)||<c(1 + |t|^3)$.

We now consider a configuration $\tI_i\in\tilde{\cal K}$,
${\bf B}_i\in {\cal F}(\tI_i)$, $\tilde{f}_i\in {\cal C}^\infty_0
(V_+)$, $\epsilon_i >0$, $i=1,\dots,n$ such that the regions
$\tI_i+tV_{\epsilon_i}(f_i)$ are mutually spacelike for
large $t$. Then the limit
\begin{equation}
 \lim_{t\rightarrow\infty}{\bf B}_n(t)\cdots {\bf B}_1(t) {\bf
 \Omega}
\end{equation}
exists and may be interpreted as a vector describing an
outgoing configuration of $n$ particles
with state vectors ${\bf \Psi}_i={\bf B}_i(t){\bf \Omega}$.
As long as the localizations $\tI_i$ are kept fixed
the scattering vectors depend only on these 1-particle vectors,
hence we may write
\begin{equation}
 \lim_{t\rightarrow\infty}{\bf B}_n(t)\cdots {\bf B}_1(t) {\bf
 \Omega}=({\bf \Psi}_n,\tI_n)\times\cdots \times ({\bf \Psi}_1,\tilde
 {I}_1).
\end{equation}
It is also easy to see how the Poincar\'e group acts and
how the scattering vectors depend on the order of one
particle vectors:
\begin{equation}
 {\bf U}(L)
 ({\bf \Psi}_n,\tI_n)\times\cdots \times ({\bf \Psi}_1,\tilde {I}_1)=
 ({\bf U}(L)
 {\bf \Psi}_n,L\tI_n)\times\cdots \times
 ({\bf U}(L) {\bf \Psi}_1,L\tilde {I}_1).
\end{equation}
and
\begin{equation}
\label{commute}
 ({\bf \Psi}_{\sigma(n)},\tI_{\sigma(n)})
 \times\cdots \times ({\bf \Psi}_{\sigma(1)},\tilde {I}_{\sigma(1)})=
 \varepsilon(b)({\bf \Psi}_n,\tI_n)
 \times\cdots \times ({\bf \Psi}_1,\tilde {I}_1),
\end{equation}
where $b$ is the cylinder braid defined in Section 2. Note that
$\varepsilon$ acts via the vacuum representation. Since the intertwiner
describing the transition from $\tI_1$ to other sheets is
trivially represented in the vacuum,
$\pi_0\circ\varepsilon$
is
actually a representation of the groupoid of colored braids
on the plane.

In order to understand the dependence of the scattering vectors
on the localizations $\tI_i\in\tilde{\cal K}$ let us assume that
there exists for some $j\in\{1,\dots,n\}$ a localization
$\tilde{J}_j\in\tilde{\cal K}$, a field operator
${\bf C}_j\in {\cal F}(\tilde{J}_j)$ and a test function
$g_j$ with ${\rm supp}\tilde{g}_j\subset m_{j} V_\epsilon(f_j)$
such that
(with ${\bf C}_j(t)$ defined in analogy to (\ref{creation}))
\begin{equation}
 {\bf \Psi}_j = {\bf C}_j(t) {\bf \Omega},
\end{equation}
and $\tilde{J}_j+tV_{\epsilon}(f_j)$ is spacelike to
$\tI_i+tV_\epsilon(f_i)$ for $i\neq j$ and for large $t$.
If $j=1$ the scattering vector does not change
when ${\bf B}_j(t)$ is replaced by
${\bf C}_j(t)$.
If $j\neq 1$
we first commute ${\bf B}_j(t)$ to the right, then replace
it by ${\bf C}_j(t)$ and commute it back to the $j$-th place.
The whole procedure amounts to an application of an intertwiner
$\varepsilon(b)$ to the scattering vector where $b$ is a pure
cylinder braid obtained by the prescription in Section 2.

We now turn to a computation of the scalar product of
scattering vectors. Let $V_i\subset H_1$ be compact and
$\tI_i\in\tilde{\cal K}$, $i=1,\dots,n$ such that for
suitable neighborhoods $V_i^\epsilon$ of $V_i$ in $V_+$ the
regions $tV_i^\epsilon+\tI_i$ are mutually spacelike
for large $t$, and let $f_i$ be test-functions with ${\rm supp}
\tilde{f}_i\subset V_i^\epsilon$.
Let ${\bf B}_i,{\bf C}_i\in \F(\tI_i)$ with associated
single particle representations
$\ro_i$ and $\sigma_i$, respectively, $i=1,\dots,n$.
Let $T\in \A(S_0)$ be an intertwiner from
$\sigma_1\cdots\sigma_n$ to $\ro_1\cdots\ro_n$.
Then, with ${\bf \Psi}_i={\bf B}_i(t){\bf \Omega}$,
${\bf \Phi}_i={\bf C}_i(t){\bf \Omega}$, we find the following

\begin{Theorem} Let $\phi_i$ be the unique left inverse of
$\ro_i$, $i=1,\dots,n$.
\begin{list}{(\roman{enumi})}{\usecounter{enumi}}
\item If $\ro_i$ is not equivalent to $\sigma_i$ for some $i\in\{1,\dots
,n\}$, then
\begin{equation}
\left(
({\bf \Psi}_n,\tI_n)\times\cdots\times ({\bf \Psi}_1,\tI_1),
T\; ({\bf \Phi}_n,\tI_n)\times\cdots\times ({\bf \Phi}_1,\tI_1)\right) =0.
\end{equation}
\item If $\ro_i=\sigma_i,\; i=1,\dots,n$, then
\begin{equation}
\left(
({\bf \Psi}_n,\tI_n)\times\cdots\times ({\bf \Psi}_1,\tI_1),
T\; ({\bf \Phi}_n,\tI_n)\times\cdots\times ({\bf \Phi}_1,\tI_1)\right)
= \phi_n\cdots\phi_1(T) \prod_i\; ({\bf \Psi_i},{\bf \Phi_i}) .
\end{equation}
\end{list}
\end{Theorem}

\noindent {\it Proof.} We apply recursively Lemma 2.2 to the left
hand side of the above expression. If $\ro_i$ is not equivalent
to $\sigma_i$ for some $i$, we obtain zero. Otherwise, we have
\begin{equation}
\chi_1\cdots\chi_n(T) \prod_i\; ({\bf \Psi_i},{\bf \Phi_i}).
\end{equation}
$\chi_1\cdots\chi_n$ is the standard right inverse on $\ro_{1}\cdots\ro_{n}$
and therefore agrees with the standard left inverse
$\phi_n\cdots\phi_1$ on local intertwiners.
\qed
\medskip

Thus the scattering vectors depend in a continuous way on the
one particle vectors. Since $\F(\tI)$ is dense in
$\Hi$ for all $\tI\in\tilde{\cal K}$ we find, by going to
the closure, all scattering states corresponding to single particle
states with prescribed momentum support.

For a single particle
representations $\ro$ with mass $m_{\ro}$,
let $\Hi_{\ro}(V)=\{{\bf \Psi}\in \Hi_{\ro},
{\rm sp}_{\bf U}{\bf \Psi}\subset m_{\ro} V\}$. Given
an irreducible endomorphism $\sigma\in\Des$,
let $\Hi_{{\bf V,\ro},\sigma}
\subset \Hi_\sigma$ denote the Hilbert space
spanned by the scattering vectors
\begin{equation}
 T\; ({\bf \Psi}_n,\tI_n)\times\cdots\times ({\bf \Psi}_1,\tI_1),
\end{equation}
where ${\bf \Psi}_i\in \Hi_{\ro_i}(V_i)$, ${\bf V}=(V_1,
\dots,V_n)$, ${\bf \ro}=(\ro_1,\dots,\ro_n)$ and
$T\in \A(S_0)$ is an intertwiner from $\ro_1\cdots\ro_n$
to $\sigma$. Finally, we define
\begin{equation}
 \Hi_{{\bf V,\ro},\sigma}^{(0)}=\left(\bigotimes_i\Hi_
 {\ro_i}(V_i)\right) \otimes \Hi_{\ro_1\cdots\ro_n,\sigma} ,
\end{equation}
where
$\Hi_{\ro_1\cdots\ro_n,\sigma}$ is the space of
intertwiners $T\in \A(S_0)$ from $\ro_1\cdots\ro_n$ to
$\sigma$, equipped with the scalar product
\begin{equation}
 (S,T)\bbbone =\phi_n\cdots\phi_1(S^*T)=TS^*\frac{d_\sigma}{
 d_{\ro_1}\cdots d_{\ro_n}},
\end{equation}
where $d_\ro$ is the statistical dimension of $\ro\in\Des$ and
$TS^*$ is a multiple of the identity as a local self-intertwiner
of the irreducible endomorphism $\sigma\in\Des$.
\medskip

Theorem~3.2 implies that for
each configuration ${\bf \tI}=(\tI_1,\dots,\tI_n), \tilde{I}_i
\in\tilde{\cal K}$ such that the sets $\tI_i+tV_i$ are
mutually spacelike for large $t$ and for each unitary local
self-intertwiner $U$ of $\ro_{1}\cdots\ro_{n}$, there exists an
isometric embedding
\begin{equation}
i({\bf \tI}, U): \Hi_{{\bf V,\ro},\sigma}^{(0)} \rightarrow \Hi_\sigma
\end{equation}
given by
\begin{equation}
\left( {\bf \Psi}_1\otimes\cdots\otimes {\bf \Psi}_n\right) \otimes T
\mapsto
T\; U \; ({\bf \Psi}_n,\tI_n)\times\cdots\times ({\bf \Psi}_1,\tI_1).
\end{equation}

Embeddings related to different choices of ${\bf \tI}$
are related by a pure braid which acts from
the right on the intertwiner $U$.
This braid can be chosen to be a local intertwiner,
as it acts via the vacuum representation (cf. (\ref{intert})).
Thus the space of scattering states
$\Hi_{{\bf V,\ro},\sigma}$ does not depend on ${\bf \tI}$ or $U$.

It is easy to see that the embeddings do not change when the
localizations are translated or made smaller. Hence we may label
the configurations ${\bf \tI}$ by points $\tilde{r}_i$ in the
covering space of the
spacelike hyperboloid $\{x\in \M,x^2=-1\}$. Moreover, the
embeddings are locally constant in $\tilde{r}_2,\dots,\tilde{r}_n$
and are globally constant in $\tilde{r}_1$.

To describe the global structure of the space of scattering states
let us introduce the following notation.
Let $e_0\in H_1$ be arbitrary.
A configuration of disjoint
particle velocities
$q_i=\frac{p_i}{m_i}\neq q_j=\frac{p_j}{m_j},i\neq j$,
is called {\it regular} (with respect to $e_0$)
if there are mutually
spacelike cones $S_{i}$ with apex $e_0$ such that
$q_{i}\in S_{i}$ (in particular
$q_{i}\neq e_0$).
To a regular configuration ${\bf q}=(q_1,\dots,q_n)$
we associate a configuration of spacelike directions
${\bf r}=(r_{1},\ldots,r_{n})$, with $r_i=q_i-e_0$.

We now pick
a regular
``reference'' configuration ${\bf q}^{0}$ and
choose $n$ homotopy classes of
paths of
spacelike cones ${\tI_{i}}, i=1,\ldots,n$
whose endpoints $e(\tI_{i})$ correspond to the canonical directions
$r_{i}^{0}$. We label these homotopy classes
by ${\bf \tilde{r}}^{0}$. We also choose a unitary local self-intertwiner
$U^{0}$. These choices then determine a reference embedding
$i({\bf \tilde{r}}^{0},U^{0})$
for a neighborhood ${\bf V}^{0}$ of ${\bf q}^{0}$.
(Note that ${\bf V}^{0}$ may contain also nonregular points.)

In the next step, we cover the configuration
space of non-coinciding velocities by embeddings around
regular
configurations ${\bf q}$. (This is possible.)
We label these embeddings
by paths $\gamma_{\bf q}$ from ${\bf q}^{0}$ to ${\bf q}$ which have
the property that none of the $n$ velocities passes through
$e_0$.
The embedding corresponding to $\gamma_{\bf q}$ is then
specified in the following way.
As before, we describe the spacelike directions ${\bf r}^{0}$ by the
$n$ points $(r_{1}^{0},1),\ldots,(r_{n}^{0},n)$ on the cylinder
$S^{1} \times \bbbr_{+}$. As long as ${\bf q}(t)$ is ordered, the
corresponding path ${\bf r}(t)$ is canonically determined.
At a critical point, where ${\bf q}(t)$ ceases to be ordered, two
directions $r_{i}$ and $r_{j}$ coincide \footnote{This is the generic
case. The general case is covered by the geometrical description
given below.}. In a neighborhood of this
critical point we define ${\bf r}(t)$ by the following prescription:
we move the direction $r_{k}$ corresponding to the {\it smaller} velocity
from $(r_{k},k)$ to $(r_{k},1/2)$, then change the two directions
past each other
and finally move $(r_{k}',1/2)$ back to $(r_{k}',k)$. Geometrically, this means
that the points ${\bf r}(t)$ on the cylinder, viewed from the
$(S^{1},0)$-end of the cylinder and looking in the long direction,
perform the same
motion as the velocities ${\bf q}(t)$ when viewed from $e_0$
(compare figure~2). We
denote the so determined path ${\bf r}(t)$ by
${\bf \gamma}^{r}=(\gamma_{1}^{r}, \ldots , \gamma_{n}^{r})$.

\begin{figure}
\hspace*{4.25cm}
\epsfxsize=8.5cm
\epsfbox{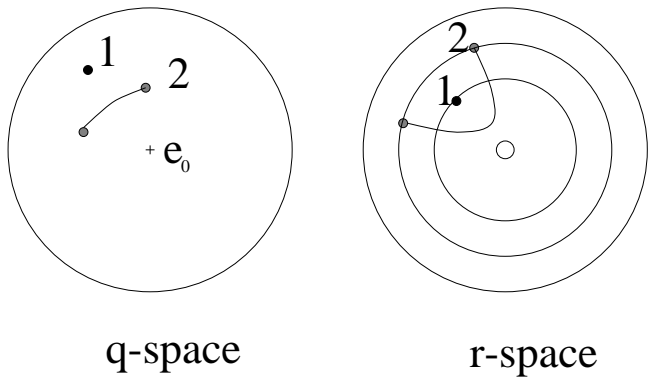}
\caption{A path ${\bf q}(t)$ and the corresponding path ${\bf r}(t)$.}
\end{figure}

Each path $\gamma_{i}^{r}$ lifts to a unique path
$\tilde{\gamma}_{i}^{r}$ from $\tilde{r}_{i}^{0}$ to
$\tilde{r}_{i}$. We denote the corresponding
configuration by
$\tilde{\bf r}=(\tilde{r}_{1}, \ldots , \tilde{r}_{n})$.
Each path $\gamma_{\bf q}$ therefore determines
a pure braid on the cylinder, namely the homotopy class
\begin{equation}
\label{pbraid}
b(\gamma_{\bf q}) = \left( \tilde{\bf r} \right)^{-1} \circ \gamma^{r} \circ
\left( \tilde{\bf r}^{0} \right).
\end{equation}
Note that $b$ defines a homomorphism of the groupoid of colored braids
on the cylinder (corresponding to paths $\gamma_{\bf q}$, where
${\bf q}$ is a permutation of ${\bf q}^{0}$) to the pure braid group
of the cylinder. This homomorphism is actually an automorphism
of the pure braid group of the cylinder when restricted to closed
paths $\gamma_{{\bf q}^{0}}$. (Both properties can be easily seen from
the geometrical description.) Note also, that the image of a given
path does not depend on the choice of $e_{0}$ locally.

\noindent The embedding corresponding to $\gamma_{\bf q}$ is
then determined by
\begin{equation}
\label{embed}
i(\gamma_{\bf q}) = i\left(\tilde{\bf r}, U^{0}\; \varepsilon(
b(\gamma_{\bf q}))^{-1} \right).
\end{equation}

Next we want to calculate transition functions between different embeddings.
Let ${\bf q}' \in {\bf V}_{1} \cap {\bf V}_{2}$, where
${\bf V}_{i}$ is the neighborhood of ${\bf q}_{i}$ for which
the corresponding embedding is well defined. In a first step we
want to calculate the transition function between embeddings
$\gamma_{1}=\gamma_{{\bf q}_{1}}$ and $\gamma_{2}=\gamma_{{\bf q}_{2}}$
where $\gamma_{2}$ is obtained from $\gamma_{1}$ by composition with
the path ${\bf q}_{1} \rightarrow {\bf q}' \rightarrow {\bf q}_{2}$ in
${\bf V}_{1}\cup {\bf V}_{2}$.
In this case we have the following lemma

\begin{Lemma}
The transition function between two such embeddings is given as
\begin{equation}
\label{trans}
i_{2} ^{-1} \circ i_{1} = id.
\end{equation}
\end{Lemma}

\noindent {\it Proof.} We have to calculate how the vector
\begin{equation}
i_{1}({\bf \Psi}_{1}\otimes\cdots\otimes {\bf \Psi}_{n} \otimes T) =
T \; U_{1}\; ({\bf \Psi}_{n},\tI^{1}_n)\times\cdots\times
({\bf \Psi}_1,\tI^{1}_1)
\end{equation}
transforms under changing the localization cones ${\bf \tI}^{1}$ to
${\bf \tI}^{2}$. Without loss of generality we can assume that
${\bf \tI}^{2}$ differs from  ${\bf \tI}^{1}$ only by the relative
orientation of two adjoining cones, as the general case can be reduced
to this case recursively.
To calculate the unitary intertwiner in this case,
we use (\ref{commute}) to
commute the single particle vector
corresponding to the {\it smaller} velocity
to the right. We
then change its localization to
the localization corresponding to ${\bf \tI}^{2}$ and commute it back,
using (\ref{commute}) again.
(We have to change the localization region of the
vector with the
smaller velocity in order to guarantee that the two localization
regions become mutually spacelike for large $t$
for both ${\bf \tI}^{1}$ and ${\bf \tI}^{2}$.)
Thus the unitary intertwiner which relates the two localizations
is just the pure braid obtained from the
path from ${\bf q}_{1}$ to ${\bf q}_{2}$ in
${\bf V}_{1}\cup {\bf V}_{2}$ via (\ref{pbraid}).
As the two embeddings differ by this braid, the transition function
(\ref{trans}) is trivial.
\qed
\medskip

Now let $\Phi\in \Hi_\sigma$ and denote the projector onto
$\Hi_{{\bf V,\ro},\sigma}$ by $E_{\bf V}$. For each $\gamma_{\bf q}$
such that $i(\gamma_{\bf q})$ is an embedding $\Hi^{(0)}_{{\bf V,\ro},\sigma}
\rightarrow \Hi_\sigma$, we can associate to $E_{\bf V} \Phi$
\begin{equation}
i(\gamma_{\bf q})^{-1} (E_{\bf V} \Phi) \in
\Hi^{(0)}_{{\bf V,\ro},\sigma}.
\end{equation}
Because of the above lemma,
\begin{equation}
i(\gamma_{\bf q})^{-1} (E_{{\bf V}\cap {\bf V}'} \Phi) =
i(\gamma_{{\bf q}'})^{-1} (E_{{\bf V}\cap {\bf V}'} \Phi)
\end{equation}
if $\gamma_{\bf q}$ and $\gamma_{{\bf q}'}$ differ by a path in
${\bf V} \cup {\bf V}'$. Thus we can extend the map
\begin{equation}
\Phi(\gamma_{\bf q}) = i(\gamma_{\bf q})^{-1} (E_{\bf V} \Phi) \in
\Hi^{(0)}_{{\bf V,\ro},\sigma} ,
\end{equation}
where ${\bf V}$ is some neighborhood of ${\bf q}$, to arbitrary
(not necessarily
regular) configurations ${\bf q}$ of noncoinciding velocities.
As we have given an explicit definition of the embeddings
(\ref{embed}), we can calculate the covariance property of
$\Phi(\gamma_{\bf q})$. We find
\begin{equation}
\label{cova}
\Phi(\gamma_{\bf q} \circ \gamma_{{\bf q}^{0}}) =
\Phi(\gamma_{\bf q}) \; U^{0} \;
\varepsilon\left( b\; (\gamma_{{\bf q}^{0}})\right)\;
\left( U^{0} \right)^{-1},
\end{equation}
where $\gamma_{{\bf q}^{0}}$ is a pure braid in the homotopy class
of ${\bf q}^{0}$, $b$ is the automorphism of the pure braid group
defined in (\ref{pbraid}),
and the right action of $\gamma_{{\bf q}^{0}}$ is the global action of
the structure
group in the universal covering space.
To prove (\ref{cova}) it is sufficient to consider
product vectors
\begin{equation}
E_{\bf V} \Phi =
T \; ({\bf \Psi}_{n},\tI_{n}) \times \cdots \times
({\bf \Psi}_{1}, \tI_{1}),
\end{equation}
where ${\bf \tI}$ corresponds to $\gamma_{\bf q}$. Then
the preimage under $i(\gamma_{\bf q} \circ \gamma_{{\bf q}^{0}})$ is
given by
\begin{equation}
{\bf \Psi}_{1} \otimes \cdots \otimes {\bf \Psi}_{n} \otimes T \;
U^{-1}_{\gamma_{\bf q}\circ \gamma_{{\bf q}^{0}}},
\end{equation}
where we have
\begin{equation}
\begin{array}{rcl}
{\displaystyle
U^{-1}_{\gamma_{\bf q}\circ \gamma_{{\bf q}^{0}}} = \left( U^{0}
\varepsilon\left(b\;(\gamma_{\bf q}\circ
\gamma_{{\bf q}^{0}})\right)^{-1} \right)
^{-1}} & = &
{\displaystyle \varepsilon\Bigl(b\; (\gamma_{\bf q})\Bigr) \;
\varepsilon\left(b\;(\gamma_{{\bf q}^{0}})\right)\;
\left( U^{0} \right)^{-1}} \\
& = & {\displaystyle
\left(U^{0} \; \varepsilon\Bigl(b\;(\gamma_{\bf q})\Bigr)^{-1}
\right)^{-1} \;
U^{0} \; \varepsilon\left(b\;(\gamma_{{\bf q}^{0}})\right)\;
\left(U^{0}\right)^{-1}}.
\end{array}
\end{equation}
This proves (\ref{cova}).
\medskip

As the braids (\ref{pbraid}) are locally independent of $e_{0}$,
$\Phi(\gamma_{\bf q})$ is locally independent of $e_{0}$. To show
that $\Phi(\gamma_{\bf q})$ is also globally independent of $e_{0}$,
it is sufficient to prove that the covariance corresponding to the
motion of any velocity around $e_{0}$ is trivially represented.
To this end let us observe that we can extend the covariance
(\ref{cova}) to the colored braid group of the cylinder
(cf.\ the above remark about $b$). Then, the fact that the motion of
the first velocity around $e_{0}$ is trivially represented already
implies that the same holds for any velocity.
\smallskip

Now, if $\Hi_{{\bf V,\ro},\sigma}$ has the structure of a function
space over the configuration space of non-coinciding velocities (this
can be shown, for example, under the assumption of Lorentz covariance
as assumed here),
$\Phi(\gamma_{\bf q})$ is a function on the universal covering
space, possessing the covariance property (\ref{cova}).
These functions are in one-to-one correspondence with sections
in the associated vector bundle. We have therefore shown that
the space of scattering vectors has the structure of
the Hilbert space of square integrable sections of
an associated vector bundle
over the configuration space of non-coinciding velocities.
This is precisely the structure proposed by Mund and Schrader in
\cite{MuSch}.

\section*{Acknowledgements}
We gratefully acknowledge helpful discussions with Karl-Henning Rehren,
Robert Schrader and Bert Schroer who gave us numerous useful hints.

M.R.G. was supported by the `Universit\"{a}t Hamburg' and by the
`Graduiertenkolleg Theoretische Elementarteilchen' while visiting
Hamburg during the final stages of this work.

\end{document}